\documentclass[twosides,twocolumn]{article}
 \usepackage[T1]{fontenc}
\usepackage{amsmath, amsthm, amsfonts}
\usepackage{setspace}
\usepackage{authblk}
\usepackage[pdfborder={0 0 0}]{hyperref}
\usepackage[utf8]{inputenc}
\usepackage[english]{babel}		
\usepackage{color}							
\usepackage{exscale}						
\usepackage{times}							
\usepackage{amssymb}						
\usepackage{graphicx}						
\usepackage{pgf,pgfarrows,pgfnodes,pgfshade}	
\usepackage{float}

\usepackage[hmarginratio=1:1,top=32mm,left=20mm,columnsep=20pt]{geometry} 
\usepackage{hyperref} 
\usepackage{geometry}
\usepackage{cases}
\usepackage{float}
\restylefloat{figure}
\usepackage{chngcntr}
\usepackage{float}
\usepackage{wrapfig}
\usepackage{graphicx}
\restylefloat{figure}
\usepackage{framed}
\usepackage[german]{varioref}
\usepackage{textcomp}
\usepackage{csvsimple}
\newcommand{\picwidthSMALL}{0.5}

\theoremstyle{definition}

\theoremstyle{remark}

\def\RR{\mathbb{R}}

\usepackage{tikz}
\DeclareRobustCommand\full  {\tikz[baseline=-0.6ex]\draw[thick] (0,0)--(0.5,0);}
\DeclareRobustCommand\dotted{\tikz[baseline=-0.6ex]\draw[thick,dotted] (0,0)--(0.54,0);}
\DeclareRobustCommand\dashed{\tikz[baseline=-0.6ex]\draw[thick,dashed] (0,0)--(0.54,0);}

\title{An extended model of wine fermentation including aromas and acids}
\author[1]{Jan Bartsch*}
\affil{University of Würzburg, Institute for Mathematics\\ 
	97974 Würzburg\\ Germany\\ \newline
\url{jan.bartsch@mathematik.uni-wuerzburg.de}}
\author[2]{Alfio Borzì}
\affil{University of Würzburg, Institute for Mathematics\\
	97974 Würzburg\\ Germany\\
\url{alfio.borzi@mathematik.uni-wuerzburg.de}}
\author[3]{Christina Schenk}
\affil{Trier University, Department of Mathematics\\
	54286 Trier\\ Germany\\
\url{christina.schenk@uni-trier.de}}
\author[4]{Dominik Schmidt}
\affil{Geisenheim University, Modeling and Systems Analysis \\
	65366 Geisenheim\\ Germany\\
\url{dominik.schmidt@hs-gm.de}}
\author[5]{Jonas Müller}
\affil{Geisenheim University, Modeling and Systems Analysis \\
	65366 Geisenheim\\ Germany\\
\url{jonas.mueller@hs-gm.de}}
\author[6]{Volker Schulz}
\affil{Trier University, Department of Mathematics\\
54286 Trier\\ Germany\\ \newline
\url{volker.schulz@uni-trier.de}}
\author[7]{Kai Velten}
\affil{Geisenheim University, Modeling and Systems Analysis\\
65366 Geisenheim\\ Germany\\
\url{kai.velten@hs-gm.de}\newline corresponding author:*}

\begin{document}


\maketitle

\abstract{The art of viticulture and the quest for making  wines has a long
tradition and it just started recently that mathematicians entered this field with their main
contribution of modelling alcoholic fermentation.
These models consist of systems of ordinary differential equations that describe the
kinetics of the bio-chemical reactions occurring in the fermentation process. \\ The aim of this paper is to present a new model of wine fermentation that accurately describes
the yeast dying component, the presence of glucose transporters, and the formation
of aromas and acids. Therefore the new model could become a valuable tool to
predict the taste of the wine and provide the starting point for an emerging control technology
that aims at improving the quality of the wine by steering a well-behaved fermentation process
that is also energetically more efficient. Results of numerical simulations are presented that successfully
confirm the validity of the proposed model by comparison with real data.}  



\section{Introduction}
The ambition for making excellent wines has a very long 
history \cite{pellechia2006wine} and focuses on the crucial step of a `perfect' alcoholic fermentation. 
For this quest, also mathematics is contributing by 
investigating models consisting of systems of ordinary differential equations (ODEs) that describe the kinetics of the bio-chemical reactions occurring in the fermentation process 
and provide the starting point for an emerging control technology that aims at improving the quality of 
the wine by steering a well-behaved fermentation process that is also energetically more efficient. 

Among some of the most representative and recent models of wine fermentation, we refer to \cite{david2010dynamical} and \cite{David}, where the evolution of the yeast biomass together with the concentrations of assimilable nitrogen, sugar, and ethanol are modeled.
In this model, the growth of the yeast population that consumes nitrogen is governed by a Michaelis-Menten term, which is often used for the description of enzymatic biological reactions \cite{michaelis1913kinetik}. 
Another kinetic component of this model governs the conversion of sugar into ethanol. This component takes into account the fact that high concentration of ethanol decelerates the fermentation of sugar, which 
corresponds to an inhibition term. 
A similar model is also presented in \cite[Chapter~3.10.2]{Velten2009}, whereas an advanced model is proposed in \cite{malherbe2004modeling} that emphasizes the effect of nitrogen and takes into account the fact that this 
substance is also needed for the sugar transport in the yeast cells. 
Further, in \cite{colombie2007modeling,Zenteno2010} thermal phenomena in the fermentation process, as the production of heat by the yeast and the heat loss to the environment, are considered. 

In application, a first attempt to formulate wine fermentation control problems can be found in \cite{sablayrolles2009control} with the aim to improve energy consumption and the aromatic profile. Further efforts to develop simulation and optimization schemes for wine fermentation can be found in \cite{CharnomordicDavidDochainHilgertMouretSablayrollesVandeWouwer2010,Goelzer2009}.

Our contribution to this research effort is manifold, i.e. reflecting the multiple objectives of the project 
ROENOBIO that focused on the modelling and optimization of the 
wine fermentation process \cite{BorziMerger2014,BorziMerger,MergerRD2017,mueller2015application,EnergOpt15,Schenk,schmidt2016numerical}. In 
this work and any other improvement in simulation and 
optimization of wine fermentation, a differential model of fermentation of increasing sophistication seems 
indispensable.That is why in this paper we aim at 
developing a new model that improves upon existing ones by including and merging together different components 
like the yeast dying behavior, glucose transporters, and the presence of aromas and acids.

Wine fermentation is a very complex bio-chemical process 
where each bio-chemical component plays an important 
role. In particular, oxygen plays a crucial role for yeast activity, helping this to cope with the initial high concentrations of sugar and nutrients. However, 
oxygen which is not consumed by the yeast leads to oxidation of the wine; see e.g. \cite{oxygenarticle}. During fermentation yeast grows by metabolizing sugar in the presence of nutrients such as nitrogen. The consumed sugar is converted into ethanol. On the other hand, alcohol inhibits yeast growth because it is toxic 
for the yeast, see \cite{JarzebskiMalinowski89,LeaoUden82,SainzPizarro03} for further references and attempts to model the influence of ethanol on the fermentation rate. In fact, the alcohol which is produced 
consists of many components of which ethanol is the most relevant one. Further, as discussed in \cite{David}, the  yeast needs certain proteins, called \emph{transporters}, for the assimilation of sugar and nitrogen.

Our work starts with the basic model proposed in \cite{David}, which governs the time evolution of the yeast $X$, assimilable nitrogen $N$, ethanol $E$ and sugar $S$. In \cite{BorziMerger2014}, this model is extended by 
including: 1) a term that describes the yeast dying phase with respect to the toxic influence of alcohols (see below); 
2) an ODE for the time evolution of oxygen $O_2$; 3) the impact of oxygen and sugar on the yeast production. 
For clarity of illustration see the terms in \textbf{bold} in (\ref{align:arXiv}) and compare with \cite{David}:

\begin{equation}
\begin{cases}
\begin{aligned}
\dfrac{\textup{d} X}{\textup{d} t}=&\mu_{max}(T) \dfrac{{N}}{K_N+N}\bf\dfrac{S}{K_{S_1}+S}\\&\bf\dfrac{O_2}{K_{O}+O_2}{X}-\boldsymbol\varphi({E}){X}\\
\dfrac{\textup{d} {N}}{\textup{d} t}=&-k_1\mu_{max}(T) \dfrac{{N}}{K_N+{N}}\bf\dfrac{{S}}{K_{S_1}+{S}}\\&\bf\dfrac{{O_2}}{K_{O}+{O_2}}{X}\\
 \dfrac{\textup{d} {E}}{\textup{d} t}=&\beta_{max}(T) \dfrac{{S}}{K_{S_2}+{S}}\dfrac{K_E(T)}{K_E(T)+{E}}{X}\\
\dfrac{\textup{d}{S}}{\textup{d} t}=&-k_2\dfrac{\textup{d} {E}}{\textup{d} t}-\\& \bf k_3 \mu_{max}(T) \dfrac{{N}}{K_N+N}\dfrac{S}{K_{S_1}+S}\\&\bf\dfrac{O_2}{K_{O}+O_2}{X}\\
\bf\dfrac{\textup{d} {O_2}}{\textup{d} t}=&\bf-k_3\mu_{max}(T) \dfrac{{N}}{K_N+{N}}\dfrac{{S}}{K_{S_1}+{S}}\\&\bf\dfrac{{O_2}}{K_{O}+{O_2}}{X}\\
\end{aligned}
\end{cases}.
\label{align:arXiv}
\end{equation}

Since the yeast is also active even if no oxygen is left, we introduce a parameter $\varepsilon > 0$, to model this 
fact. However, in order to ensure that the yeast reaches a stationary state, we add an additional dying term in the equation for the yeast given 
by $-k_d\, X$; see \cite{Schenk}. These additions result in the model \textbf{(A)}.

In the following, the parameters $k_i$ represent the  \textit{stoichiometric coefficients}, which describe the substances amount ration in a chemical reaction, e.g. $k_1$ describes, in which ratio nitrogen ($N$) is used in the conversion to yeast ($X$). To have mathematically more compact equations, we often combine several stoichiometric coefficients in to one. \\

In every equation, the \emph{Michaelis-Menten-kinetics} with Michaelis-Menten constants are used to model the conversion of the substances into each other, a well known model for enzyme kinetics. The Michaelis-Menten constants are given by the substrate concentration for which the reaction rate is the half of the maximal reaction rate. Their values are presented in Table \ref{tab:Michaelis-Menten}. \\ 
The time dependent functions $\mu_{max}(\cdot),~\beta_{max}(\cdot)$ describe the specific maximal reaction rate for the given substrate, depending on the temperature.
$\varphi(\cdot):\RR^+_0 \rightarrow \RR^+$ is the function, which is used to model the toxic influence of alcohol on yeast. 
We define $\Sigma(t)$ as the sum of all different alcohols contained in wine; in our model $\Sigma(t):= E(t)+A(t)+B(t)+P(t)$. Where $A$ (isoamyl alcohol), $B$ (isobutanol), $P$ (propanol) are additional alcohols which are described in detail in Section \ref{ModelWithAromas}. \\

The toxicity function $\varphi(\Sigma)$ was proposed in 
\cite{BorziMerger2014}, based on heuristic consideration and results of measurement of wine fermentation. The 
same data was used in \cite{BorziMerger} with a new 
function identification scheme confirming 
that the following functional structure of $\varphi$ 
appropriately describes the effect of alcohol toxicity in the 
evolution of the yeast. We have 
\begin{small}
\begin{align}
\label{phifunction}
\varphi(\Sigma) = \left( 0.5 + \frac{1}{\pi}\arctan(k_{d1}(\Sigma-tol)) \right)k_{d2}(\Sigma-tol)^2 ,
\end{align}
\end{small}
where $tol$ describes the tolerance bound of the alcohol concentration and $k_{d1},k_{d2}$ are parameters, which correlate with the dying of the yeast cells related to the exceeding of the $tol$ threshold.

\subsection{Glucose transporters}
We follow the work in \cite{David} and consider the 
presence of glucose transporters that are essential 
for the assimilation of sugar (here glucose) and nitrogen. 
In fact, these transporters are responsible for the glucose passing the yeast cell membrane. Therefore 
we distinguish two nitrogen components: there exists a part of nitrogen $N_x$, which is converted into yeast $X$, and one part $N_{Tr}$, which is responsible for the synthesis of the transporters $Tr$. These components 
are included in our model in the way proposed by \cite{David} and as illustrated in the system \textbf{(B)}
as follows 
$$
\textbf{(B)}
\begin{cases}

\frac{\textup{d} Tr}{\textup{d} t} = \eta_{max}(T)\frac{N_{tr}}{N_{tr} + N_{tr}}\frac{S}{K_{S_1}+S}\frac{O_2}{K_O+O_2}X \\

\frac{\textup{d} N_{tr}}{\textup{d} t} = -k_1^{'} \eta_{max}(T)\frac{N_{tr}}{K_{tr} + N_{tr}}\frac{S}{K_{S_1}+S}\frac{O_2}{K_O+O_2}X 

\end{cases},
$$ 
where $\eta_{max}$ is given in \eqref{align:temperaturfunktionen}.
Hence, we arrive at the combined system \textbf{(A)}-\textbf{(B)}, where \textbf{(A)} also contains the glucose transporters that influence the ethanol production through the conversion of sugar. Specifically, this influence is implemented by the additional factor $(1+\Phi(T)Tr)$, where $\Phi(T)$ is a function of temperature $T$ and is given in (\ref{align:temperaturfunktionen}). 
$$
\textbf{(A)}
\begin{cases}

\frac{\textup{d} X}{\textup{d} t} = \mu_{max}(T)\frac{N_x}{K_x + N_x}\frac{S}{K_{S_1}+S}\left(\frac{O_2}{K_O+O_2}+\varepsilon\right)X
\\~~~~~~~ - k_d X -\varphi(\Sigma)X \\

\frac{\textup{d} N_x}{\textup{d} t} = -k_1 \mu_{max}(T)\frac{N_x}{K_x + N_x}\frac{S}{K_{S_1}+S}\left(\frac{O_2}{K_O+O_2}+\varepsilon\right)X \\

\frac{\textup{d} S}{\textup{d} t} = -k_2\frac{\textup{d} E}{\textup{d} t}
\\~~~~~~~ - k_3\mu_{max}(T)\frac{N_x}{K_x X + N_x}\frac{S}{K_{S_1}+S}\left(\frac{O_2}{K_O+O_2}+\varepsilon\right)X \\

\frac{\textup{d} O_2}{\textup{d} t} -k_4 \mu_{max}(T)\frac{N_x}{K_x X + N_x}\frac{S}{K_{S_1}+S}\frac{O_2}{K_O+O_2}X \\

\frac{\textup{d} E}{\textup{d} t} = \beta_{max}(T)\frac{S}{K_{S_2} + S}\frac{K_E(T)}{K_E(T) + E}X(1+\Phi(T)Tr) \\ 

\end{cases}
$$
Where the sugar-related Michaelis-Menten constant $K_1$, $K_2$ are associated to part of sugar utilized for nutrition of yeast, respectively to part required for the metabolization into alcohol.

\section{Model including aromas}
\label{ModelWithAromas}
The next important step in the development of our 
model is to take into account the presence of aroma 
compounds, which are responsible for the taste of wine; see e.g. \cite{TonesImperfections}. We consider three secondary aromas as in \cite{Rankine67,GeisenheimMaster}: propanol $(P)$, isoamyl alcohol ($A$) and isobutanol $(B)$. The ODE equations that describe the dynamics of these substances are given in \textbf{(C)} as follows

$$
\textbf{(C)}
\begin{cases}

\frac{\textup{d} P}{\textup{d} t} = k_5\mu_{max}(T)\frac{N_x}{K_x X + N_x}\frac{S}{K_{S_1}+S}\left(\frac{O_2}{K_O+O_2}+\varepsilon\right)X \\

\frac{\textup{d} A}{\textup{d} t} = k_6\beta_{max}(T)\frac{S}{K_{S_3} + S}\frac{K_E(T)}{K_E(T) + E}X(1+\Phi(T)Tr) \\  ~~~~~~+k_7\mu_{max}(T)\frac{N_x}{K_x X + N_x}\frac{S}{K_{S_1}+S}\left(\frac{O_2}{K_O+O_2}+\varepsilon\right)X \\

\frac{\textup{d} B}{\textup{d} t} = k_8\beta_{max}(T)\frac{S}{K_{S_4} + S}\frac{K_E(T)}{K_E(T) + E}X(1+\Phi(T)Tr) \\ ~~~~+  k_9\mu_{max}(T)\frac{N_x}{K_x X + N_x}\frac{S}{K_{S_1}+S}\left(\frac{O_2}{K_O+O_2}+\varepsilon\right)X \\

\end{cases}.
$$

These equations represent an extension of the 
model discussed in \cite{GeisenheimMaster} to take 
into account the coupling with the other components 
of our model. Since the propanol production stops if the assimilable nitrogen part is depleted, in the related equation only the nitrogen part $N_x$ appears; see \cite{MouretHigherAlcohols} for a related discussion. In contrast to that, there is an additional influence of sugar to isoamyl alcohol and isobutanol. Therefore, a supplementary summand has to be taken into account. 
The additional Michaelis-Menten constants related to sugar $K_{S_3}$, $K_{S_4}$ describe the saturation constants associated to the corresponding alcohols. \\
Now, to conclude the discussion on our new model 
and before we address the presence of acids, we 
give in Table \ref{tab:Parametervalues} and 
Table \ref{tab:Michaelis-Menten} the values of the parameters appearing in our \textbf{(A)}-\textbf{(B)}-\textbf{(C)} model. In Table \ref{tab:Parametervalues}, the stoichiometric coefficients are given in terms of relative ratios; in Table \ref{tab:temperaturDependence}, parameters for the functions that represent the time dependence of the reaction time are presented; in Table \ref{tab:Michaelis-Menten}, we give the Michaelis-Menten constants in $g/l$. We take the parameter values as they are given in \cite{BorziMerger2014,GeisenheimMaster,David}.

\begin{table}[H]
\centering
\begin{tabular}{|l|l| l |l|l|} \hline
  \textbf{symbol} & \textbf{value} & & \textbf{symbol} & \textbf{value} \\ \hline
 $k_1$ &  0.0115 &   & $k_1'$ & 0.160 \\ \hline
   $k_2$  &  1.9569 & & $k_3$ & 2.8424\\ \hline
 $k_4$ & 0.0004 &  & $k_5$& 0.025\\ \hline
  $k_6$ &   0.0023 & & $k_7$& 0.0005\\ \hline
  $k_8$ & 0.009 & & $k_9$& 0.0005\\ \hline
  $k_{d1}$ & 9.9676 & & $k_{d2}$& 0.0004\\ \hline
\end{tabular}
\caption{\textbf{(A)}-\textbf{(B)}-\textbf{(C)} Stoichiometric coefficients}
\label{tab:Parametervalues}
\end{table}

\begin{table}
\centering
\begin{tabular}{|l|l| l |l|l|} \hline
  \textbf{symbol} & \textbf{value} & & \textbf{symbol} & \textbf{value} \\ \hline
  $tol~~~(g/l)$ & 70 & & $\beta_1~~~((d^\circ\text{C})^{-1})$ & 0.1\\ \hline
  $\beta_2~~~(d^{-1})$ & 0.0728 & & $\mu_11~~~((d^\circ\text{C})^{-1})$ & 0.02 \\ \hline
  $\mu_2~~~(d^{-1})$ & 0 & & $\Phi_1~~~(l/(g^\circ\text{C}))$ & 0.0364 \\ \hline
  $\Phi_2~~~(l/g)$ & 0.3945  & & $\eta_1~~~((d^\circ\text{C})^{-1})$ & 0.0125833 \\ \hline
  $\eta_2~~~(d^{-1})$ & 0.1655 & & &    \\ \hline
\end{tabular}
\caption{\textbf{(A)}-\textbf{(B)}-\textbf{(C)} Values for temperature dependence; units are specified in brackets}
\label{tab:temperaturDependence}
\end{table}

\begin{table}
\centering
\begin{tabular}{|l|l| l |l|l|} \hline
  \textbf{symbol} & \textbf{values} & & \textbf{symbol} & \textbf{values} \\ \hline
 $K_X$ &  0.0007 &   & $K_{S_1}$ & 53.2669 \\ \hline
   $K_{S_2}$  &  0.1599 & & $K_{E_1}~~~(g/(l\cdot min))$ & 1.5693\\ \hline
 $K_{E_2}$ & 45.1692 &  & $K_O$& 0.0002\\ \hline
  $K_{S_3}$ & 32 & & $K_{S_4}$& 220\\ \hline
  $K_{Tr}$ & 0.174 & &$K_{AA}$ &6 \\ \hline
  $K_{MA}$ & 1.5 & &$K_{TA}$ &6  \\ \hline
 $K_{S_5}$ & 53.2669 &   &  &  \\ \hline
\end{tabular}
\caption{Michaelis-Menten constants; the unit is $g/l$ unless noted otherwise }
\label{tab:Michaelis-Menten}
\end{table}

These values also enter in the functional dependence 
on the temperature of some constants appearing in our model. This functional dependence is given in 
\eqref{align:temperaturfunktionen}. 
\begin{align}
\label{align:temperaturfunktionen}
\beta_{max}(T) = \beta_1 T - \beta_2 \notag \\
\mu_{max}(T) = \mu_1 T - \mu_2 \notag \\
\Phi(T) = \Phi_1 T - \Phi_2 \\
\eta_{max}(T) = \eta_1 T - \eta_2 \notag \\
K_E(T) = -K_{E1}T+K_{E2} \notag
\end{align}
Notice that in \eqref{align:temperaturfunktionen}, the relation between temperature and maximum rate of conversion of the substances is approximated by linear functions.
 We ensure that all these functions are non-negative by projection on $\mathbb{R}^+$.
We consider linear dependence in order to be consistent with the references, and mention that it is also common to model the temperature dependence in chemical reactions with the so called \emph{Arrhenius} equation, see \cite{Laidler84}. Note that all kinetic constants depend on temperature, but we focused on the for our study most important ones. \\  
 In the simulation results presented in the following Section \ref{subsec:Simulations}, we consider that the working temperature is 18\textcelsius~in the first half of the simulation and 30\textcelsius~in the second one; see \cite{David}.

\subsection{Simulations}
\label{subsec:Simulations}
In this section, we present results of simulation 
with our \textbf{(A)}-\textbf{(B)}-\textbf{(C)} model. 
For this purpose, we use the initial values 
given in Table \ref{tab:Startvalues}. 
\begin{table}
\centering
\begin{tabular}{|l|l| l |l|l|} \hline
  \textbf{substance} & \textbf{values (g/l)} & & \textbf{substance} & \textbf{values (g/l)} \\ \hline
 $X$ & 0.2   & & $N_x$ & 0.24\\ \hline
 $S$ & 213.4 & & $O_2$ & 0.005\\ \hline
 $E$ & 0     & & $P$   & 0\\ \hline
 $A$ & 0 & & $B$& 0\\ \hline
  $Tr$ &   0 & & $N_{Tr}$& 0.22\\ \hline
  $AA$ & 0.2 & & $MA$& 5.9\\ \hline
  $TA$ & 7.3 & & &\\ \hline
\end{tabular}
\caption{Initial values}
\label{tab:Startvalues}
\end{table}

\begin{figure}
\includegraphics[width=\picwidthSMALL\textwidth]{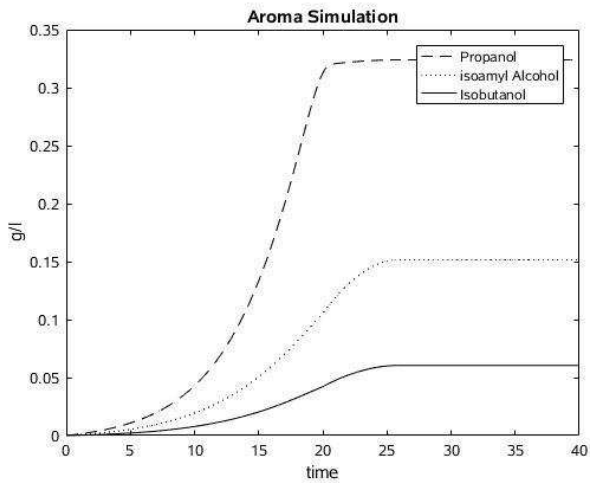}
\caption{Time evolution of aromas}
\label{fig:Aroma}
\end{figure}

\begin{figure}[H]
\includegraphics[width=\picwidthSMALL\textwidth]{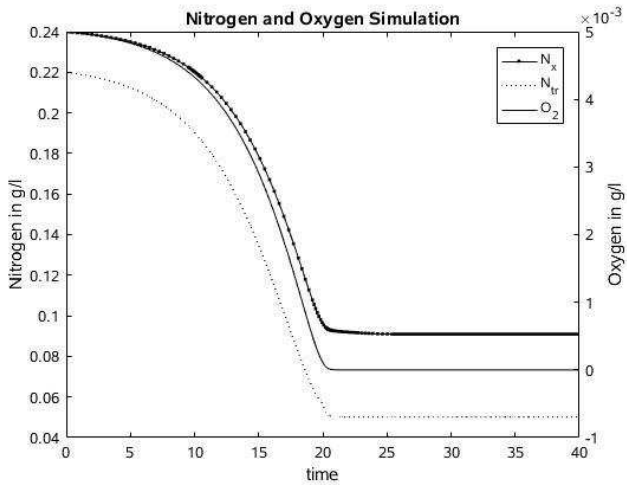}
\caption{Time evolution of nitrogen and oxygen}
\label{fig:NandO2}
\end{figure}

\begin{figure}
\includegraphics[width=\picwidthSMALL\textwidth]{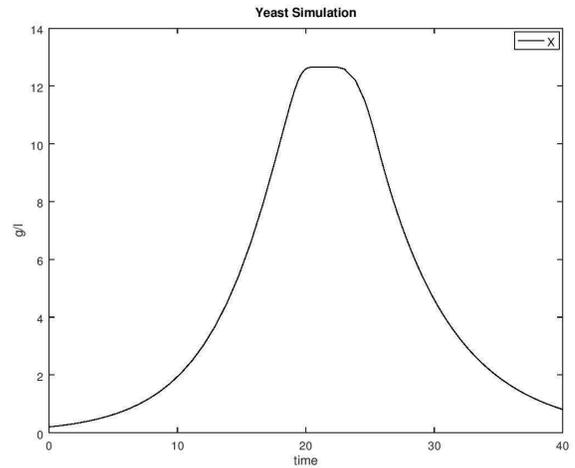}
\caption{Time evolution of yeast}
\label{fig:Yeast}
\end{figure}

\begin{figure}
\includegraphics[width=\picwidthSMALL\textwidth]{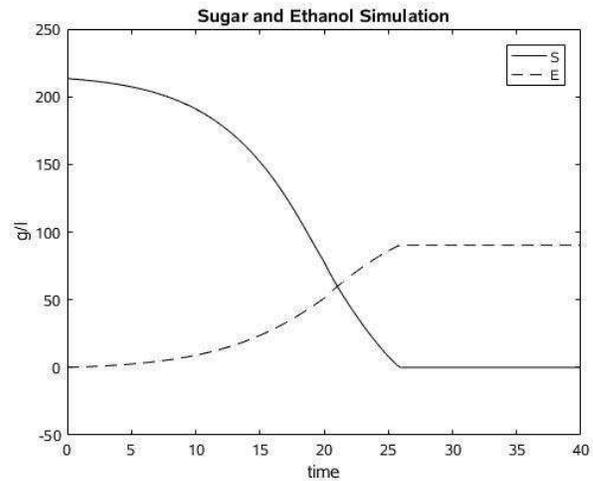}
\caption{Time evolution of sugar and ethanol}
\label{fig:SandE}
\end{figure}

\begin{figure}
\includegraphics[width=\picwidthSMALL\textwidth]{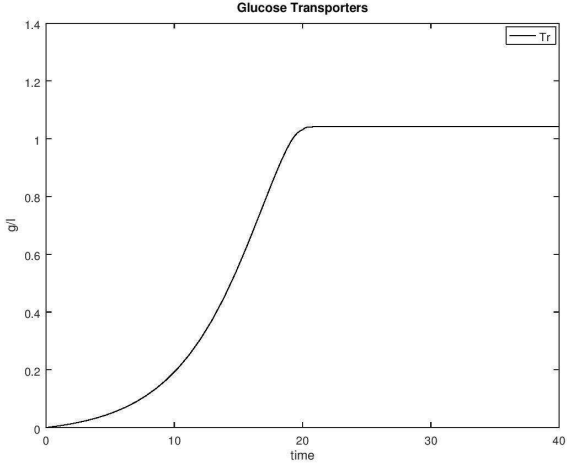}
\caption{Time evolution of glucose transporters}
\label{fig:Tr}
\end{figure}

For the numerical solution of our non-linear ODE 
system we use the Runge-Kutta method (as 
implemented by the MATLAB function \texttt{ode45}). 
We chose a time interval of 40 days, which is 
typical for a complete fermentation process \cite{David}.  In Figure \ref{fig:Aroma}-\ref{fig:Tr}, the development of the concentration of mentioned substances is given. The $y$-axis always represents this concentration in $g/l$, while on the $x$-axis the time evolution in days is given. \\
In Figure \ref{fig:Aroma}, we show the time behaviour of propanol (\dashed), isoamyl alcohol (\dotted) and isobutanol (\full). Propanol reaches the highest and isobutanol the lowest concentration at the final time. We also remark that the propanol production reaches the static stage earlier than the production of isoamyl alcohol and isobutanol due to a different dependence of propanol to sugar and nitrogen.

In 
Figure \ref{fig:NandO2}, we depict the evolution of nitrogen $N_X$ (solid-dotted) and $N_{Tr}$ (\dotted) and oxygen. In this case the solid thick line (\full) represents the oxygen concentration, and the right $y$-axis gives the concentration amount of it. We see that the depletion of oxygen is very much related to the consumption of nitrogen.

In Figure \ref{fig:Yeast}, we plot the concentration of the yeast cells, showing the toxic influence of ethanol and of the other alcohols contributing to the dying phase of yeast. In Figure \ref{fig:SandE}, we show the relation between ethanol (\dashed) and sugar (\full). One can see how ethanol is produced (by yeast) as far as sugar is available; the production of ethanol stops at the time when the sugar is depleted. In Figure \ref{fig:Tr}, we present the time evolution of the concentration of the glucose transporters in the wine.

\section{A model including acids}
The results presented in the previous section meet 
the expectation and experience of practitioners and 
encourage us to make a further step of sophistication 
of our wine fermentation model considering the 
production of acids. This addition is also useful for 
validating our model with results of measurements 
focusing on the particular acid content of the wine. 

Now, we discuss the functional dependence of 
the most relevant acids on the remaining components 
participating the wine fermentation process. 
We start discussing the \emph{acetic acid (AA)}. This 
is a by-product of alcoholic fermentation, which 
is produced by the yeast in the order of several hundreds of milligrams per liter; see \cite{C}, \cite{D}. Higher concentrations (e.g. $>1g/l$) in the final product might indicate bacterial activity during or after fermentation. 
This acid belongs to the family of \emph{volatile acids} and is responsible for acescence if its concentration reaches more than 1 g/l. Furthermore, for a good wine it must not exceed a certain concentration due to an EU Regulation, which differs depending on the particular wine produced  (see Appendix I C in \cite{EUVerordnung}). Another two important acids contained in wine are \emph{malic acid (MA)} and \emph{tartaric acid (TA)}, which we also discuss below. For detailed information concerning these acids see  \cite{OxfordWeinlexikon}.

\subsection{Acetic acid}
During fermentation, sugar conversion by lactic acid bacteria \cite{Laimburg} (otherwise called wild yeast) can lead to an increase of acetic acid concentration,
while post-fermentation spoilage can be caused by the presence of acetic acid bacteria that form
acetic acid by oxidation of ethanol (\cite{A},\cite{B},\cite{C},\cite{D}). Moreover, it is 
reasonable to assume a linear precipitation rate of the acetic acid. Further, we have an additional Michaelis-Menten constant $K_{S_5}$ describing the saturating sugar concentration for this acid. These considerations are modelled in the following equation \eqref{align:AAODE}. We have

\begin{small}
\begin{align}
 	\label{align:AAODE}
 	\frac{\textup{d} AA}{\textup{d} t} = k_{10}\theta_{max}(T)\mu_{max}(T)\frac{N_x}{K_x+ N_x}\frac{S}{K_{S_5}+S}\frac{O_2}{K_O+O_2}X \notag
  \\ +k_{11}\zeta_{max}(T)\frac{\textup{d} E}{\textup{d} t}- k_a AA.<wh
 	\end{align}
\end{small} 

\subsection{Malic acid}
Notice that malic and tartaric acids are present in the grape. The more matured the grape is, the less is the concentration of malic acid, which is responsible for the sour taste of fruits. During the fermentation process, the concentrations of these acids decreases. In the case of malic acid, this happens through \emph{malolactic fermentation}, that is, the conversion of malic acid into the softer lactic acid and $CO_2$; see \cite{OxfordWeinlexikon}. We remark that the bacteria 
that are responsible for this process need a low concentration of oxygen to act optimally. Furthermore we consider a linear precipitation rate $k_m$ of the acid. Hence, 
we arrive at the following equation \eqref{align:MAODE} describing the dynamics of malic acid:
\begin{small}
\begin{align}
\frac{\textup{d} MA}{\textup{d} t} = - k_{12} \gamma_{max}(T) \frac{MA}{K_{MA}+MA}\left(
\frac{O_2}{K_O+O_2}+k_{m}\right)
\label{align:MAODE}
\end{align} 
\end{small}

\subsubsection{Tartaric acid}
The concentration of tartaric acid is reduced through \emph{precipitation} of wine scale. This happens if the acid is no more resolvable in the developing wine; \cite{OxfordWeinlexikon}. Factors 
that contribute to the reduction of tartaric acid are low temperature and rising alcohol concentration. Therefore we model the tartaric acid dynamics by the following equation, where $k_t$ is the rate of precipitation:
\begin{small}
\begin{align}
\frac{\textup{d} TA}{\textup{d} t} = - k_{13} \gamma_{max}(T) \frac{TA}{K_{TA}+TA}\left(
\frac{K_E(T)}{K_E(T)+E}+k_{t}\right) ,
\label{align:TAODE}
\end{align} 
\end{small}
where the temperature dependent factors are again approximated by linear functions as follows 
\begin{align*}
\gamma_{max}(T) = \gamma_1 T - \gamma_2,~\tilde{\gamma}_{max}(T) = \tilde{\gamma}_1 T - \tilde{\gamma}_2 \\
\theta_{max}(T) = \theta_1 T + \theta_2,~\zeta_{max}(T) = \zeta_1 T + \zeta_2 .
\end{align*}

Summarizing, we can augment our model with the 
following equations
$$
\textbf{(D)}
\begin{cases}
\frac{\textup{d} AA}{\textup{d} t} = k_{10}\theta_{max}(T) \mu_{max}(T)\frac{N_x}{K_x + N_x}\frac{S}{K_{S_1}+S}\frac{O_2}{K_O+O_2}X
\\~~~~~~~+ k_{11}\zeta_{max}(T) \frac{\textup{d} E}{\textup{d} t} - k_a AA  \\
\frac{\textup{d} MA}{\textup{d} t} = -k_{12} \gamma_{max}(T) \frac{MA}{K_{MA}+MA}\left(\frac{O_2}{K_O+O_2}+k_m \right) \\
\frac{\textup{d} TA}{\textup{d} t} = -k_{13} \tilde{\gamma}_{max}(T) \frac{TA}{K_{TA}+TA}\left(\frac{O_2}{K_O+O_2}+k_t \right) .
\end{cases}
$$

\section{Simulation and comparison with measurements}
\begin{figure}
\includegraphics[width=\picwidthSMALL\textwidth]{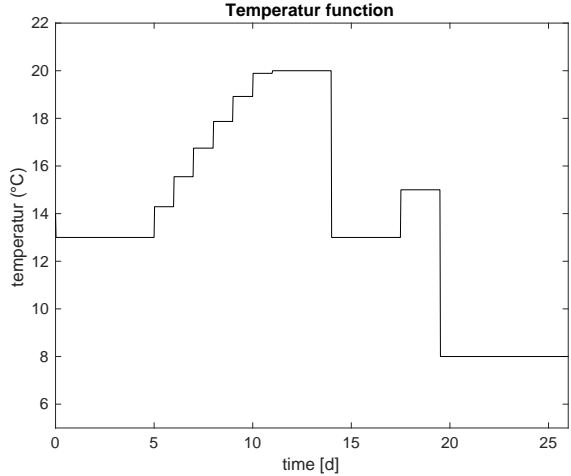}
\caption{Temperature profile in the experiment. }
\label{fig:temperature}
\end{figure}

In this section, we present results of simulation and 
comparison to real data to validate our 
new model \textbf{(A)}-\textbf{(B)}-\textbf{(C)}-\textbf{(D)}. We refer to measurements performed 
in an experiment as part of the ROENOBIO project 
in collaboration with the Dienstleistungszentrum L\"andlicher Raum (DLR) Mosel (Germany). In a time interval of almost 30 days, periodic measurements of concentrations of various important substances were taken (for more information see \cite{Schenk}).  Notice that in this 
experiment, the temperature was controlled in a way 
suggested by an independent control strategy \cite{Schenk}. This temperature profile is 
depicted in Figure \ref{fig:temperature}. 

In the following plots, we compare the results of 
simulation with our model with the measured data. 
The initial values for the simulation are 
given in Table \ref{tab:Startvalues} and the 
values of the parameters for the acetic model are given in Table \ref{tab:AcidParameters} and Table \ref{tab:AcidParameters2}.
\begin{table}
\centering
\begin{tabular}{|l|l| l |l|l|} 
 	\hline 
 	 \textbf{symbol} & \textbf{values} & & \textbf{symbol} & \textbf{values} \\
 	\hline 
 	$k_{10}$ & 0.26432 & &$k_{11}$ & 0.0015 \\ 
 	\hline 
 	$k_a$ & 0.0232 & & $k_{12}$ & 0.2994 \\
 	\hline
 	$k_m$ & 0.151 & & $k_{13}$ & 0.745 \\
 	\hline
 	$k_t$ & 0.17 & &  &  \\
 	\hline 
 	\end{tabular}
 	\caption{Stoichiometric coefficients for the acid model \textbf{(D)}.}
 	\label{tab:AcidParameters}
\end{table}

\begin{table}
\centering
\begin{tabular}{|l|l| l |l|l|} 
 	\hline 
 	 \textbf{symbol} & \textbf{values} & & \textbf{symbol} & \textbf{values} \\
 	\hline 
 	$\gamma_2$ & 1 & & $\gamma_1$ & 0.125 \\
 	\hline
 	$\tilde{\gamma}_2$ &  1 & & $\tilde{\gamma}_1$& 0.125  \\
 	\hline
	$\zeta_2$ &  1  & & $\zeta_1$&0  \\
 	\hline
 	$\theta_1$ & 0.1 & &$\theta_2$ & 0.03 \\
 	\hline
 	\end{tabular}
 	\caption{Parameters' values for the acid model \textbf{(D)} in $g/l$.}
 	\label{tab:AcidParameters2}
\end{table}

\begin{figure}[H]
\includegraphics[width=\picwidthSMALL\textwidth]{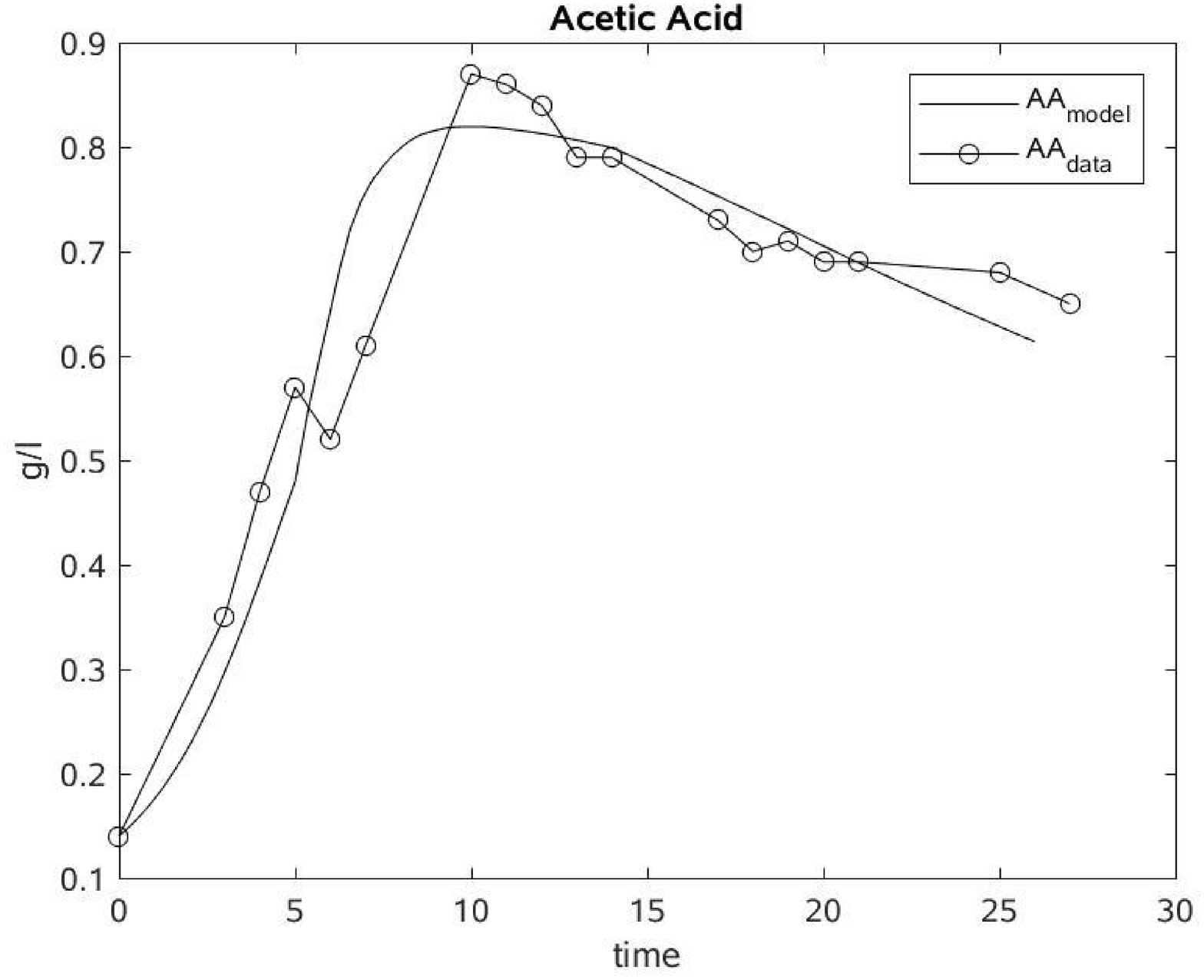}
\caption{Time development of acetic acid.}
\label{fig:AA}
\end{figure}

\begin{figure}[H]
\includegraphics[width=\picwidthSMALL\textwidth]{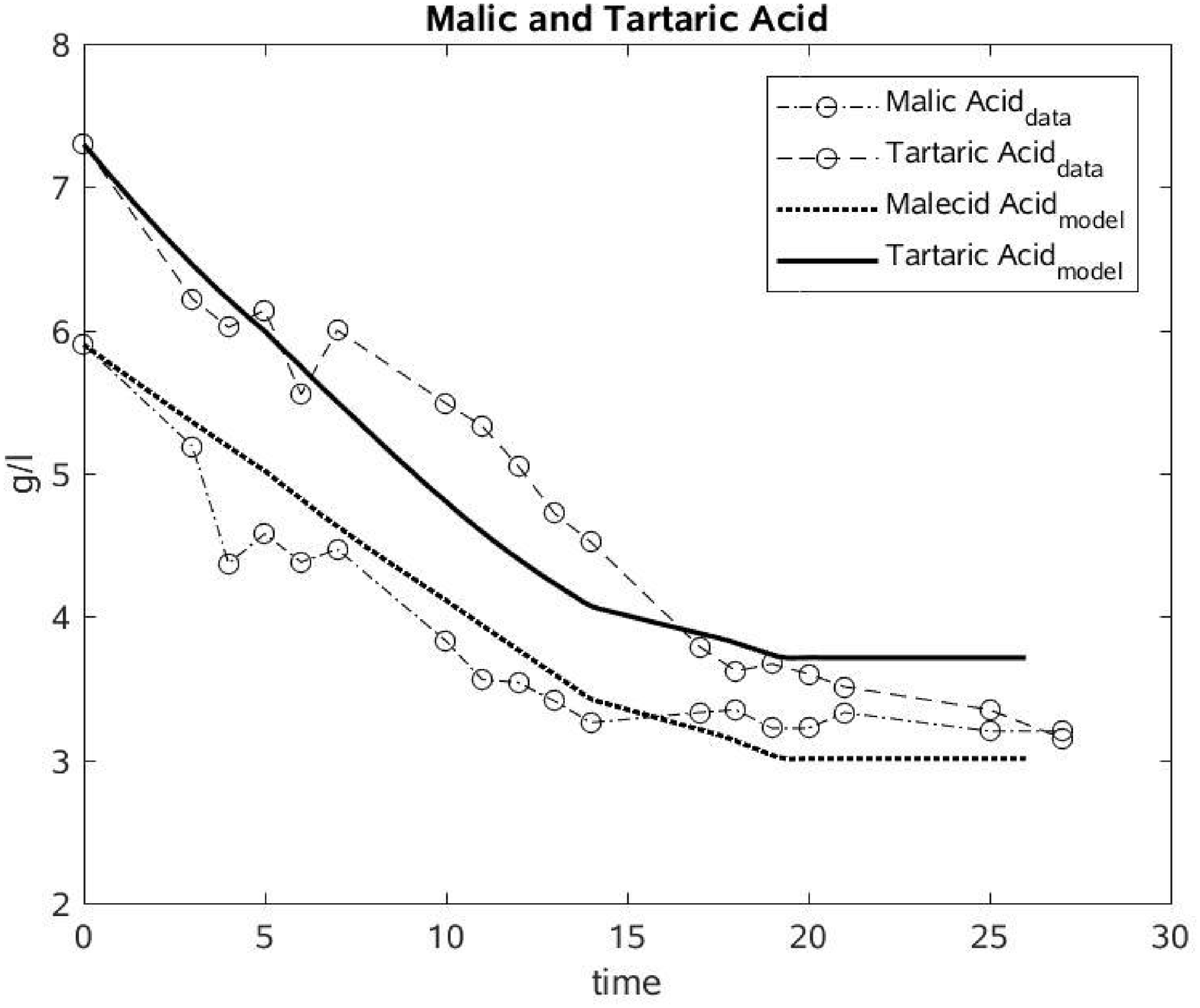}
\caption{Time development of malic and tartaric acid.}
\label{fig:MAnTA}
\end{figure}

We choose the parameters of Table \ref{tab:AcidParameters} and Table \ref{tab:AcidParameters2} based on experience. Another possibility would be to solve a parameter fitting problem using for example \texttt{MATLAB} built-in functions as \texttt{fminsearch} in combination with an ODE solver.
In Figure \ref{fig:AA}, we see measured concentration of acetic acid in $g/l$ over a time period of about 30 days. The solid line shows the result of our simulation with our extended model. In Figure \ref{fig:MAnTA}, one can see the results of the measurements of the other two chosen acids collected in the same experiment. One can see that the time evolution of all three acids is reproduced in our simulation with a striking accuracy. The acetic acid concentration raises in both experiment and simulation until a specific time and then diminishes with the same rate. In the evolution of malic (\dotted) and tartaric (\full) acid, we see different stages of decrease in the measured data, as well as in our model.

\section{Conclusion}
In this paper, a new extended model of wine fermentation has been presented. It extends previous models by including the time development of aromas and acids and is augmented by the presence of glucose transporters. Results of numerical experiments and comparison to real data have been illustrated that successfully validate the proposed model.\\
This work was done in the context of the project ROENOBIO aiming at the development of multi-dimensional simulation and control tools for an improved production and quality of white wines. The proposed model represents the core component (kinetics) of bio-chemical reaction fluid models for simulation of wine fermentation in tanks and in related control problems where an optimal fermentation pattern is driven by temperature regulation.

\section{Acknowledgements}
This work was supported by the BMBF (German Federal Ministry of Education and Research) project ROENOBIO (Robust energy optimization of fermentation processes for the production of biogas and wine) with contract number \texttt{05M2013UTA}. We would like to thank Achim Rosch and Christian von Walbrunn for providing the measurements by which we validated our results and Peter Fürst and Peter Petter (fp-sensor systems) for equipping those experiments
with their technology.


\bibliographystyle{alpha}

\newcommand{\etalchar}[1]{$^{#1}$}


\end{document}